\documentclass[superscriptaddress,twocolumn,floatfix]{revtex4-2}
\usepackage{graphicx}
\usepackage{bbm}
\usepackage{mathtools}
\usepackage{amssymb,amsmath,amsfonts,dsfont}
\usepackage{color}
\usepackage{adjustbox}
\usepackage{tabularx}
\usepackage{array}
\usepackage{multirow}
\usepackage{tabu}
\usepackage{tikz}
\usepackage{makecell}%To keep spacing of text in tables
\usepackage{url}
\usepackage{booktabs}
\usepackage{xcolor}
\usepackage{hyperref}       % hyperlinks
\usepackage{nicefrac}       % compact symbols for 1/2, etc.
\usepackage{microtype}      % microtypography
\usepackage{lipsum}
\graphicspath{{media/}}     % organize your images and other figures under media/ folder
\usepackage[braket, qm]{qcircuit} % quantum circuit
\renewcommand{\vec}[1]{\boldsymbol{#1}} 
  
\begin{document}
%% Title
\title{Variational quantum one-class classifier}

\author{Gunhee Park}
\affiliation{Division of Engineering and Applied Science, California Institute of Technology, Pasadena, CA 91125, USA}

\author{Joonsuk Huh}
\email{joonsukhuh@gmail.com}
\affiliation{Department of Chemistry, Sungkyunkwan University, Suwon, 16419, Republic of Korea}
\affiliation{SKKU Advanced Institute of Nanotechnology, Sungkyunkwan University, Suwon, 16419, Republic of Korea
}
\affiliation{
Institute of Quantum Biophysics, Sungkyunkwan University, Suwon, 16419, Republic of Korea
}

\author{Daniel K. Park}
\email{dkd.park@yonsei.ac.kr}
\affiliation{Department of Applied Statistics, Yonsei University, Seoul, 03722, Republic of Korea}
\affiliation{Department of Statistics and Data Science, Yonsei University, Seoul, 03722, Republic of Korea}

\begin{abstract}
One-class classification is a fundamental problem in pattern recognition with a wide range of applications. This work presents a semi-supervised quantum machine learning algorithm for such a problem, which we call a variational quantum one-class classifier (VQOCC). The algorithm is suitable for noisy intermediate-scale quantum computing because the VQOCC trains a fully-parameterized quantum autoencoder with a normal dataset and does not require decoding. The performance of the VQOCC is compared with that of the one-class support vector machine (OC-SVM), the kernel principal component analysis (PCA), and the deep convolutional autoencoder (DCAE) using handwritten digit and Fashion-MNIST datasets. The numerical experiment examined various structures of VQOCC by varying data encoding, the number of parameterized quantum circuit layers, and the size of the latent feature space. The benchmark shows that the classification performance of VQOCC is comparable to that of OC-SVM and PCA, although the number of model parameters grows only logarithmically with the data size. The quantum algorithm outperformed DCAE in most cases under similar training conditions. Therefore, our algorithm constitutes an extremely compact and effective machine learning model for one-class classification.
\end{abstract}
\maketitle

\section{Introduction}
\label{sec:intro}
With the growing demand for efficient and effective methods to extract useful knowledge from data, Quantum Machine Learning (QML) has emerged as a promising application of quantum technology~\cite{wittek_book,QML_book}. Many pattern recognition problems in data science can be formulated as a classification problem, which can be addressed via supervised machine learning. Several theoretical works showed that QML can be advantageous for classification in terms of runtime~\cite{PhysRevLett.113.130503,qPCA,PhysRevA.97.042315,PhysRevA.94.022342,grant_hierarchical_2018,cong_quantum_2019,Liu2021_rigorousrobust}, trainability and model capacity~\cite{Havlicek2019,abbas_power_2021}, and prediction accuracy~\cite{hur2021quantum}.

While the majority of existing works on QML for classification addresses binary problems, this work focuses on one-class classification (OCC).
One-class classification has a wide range of applications, such as anomaly detection in finance~\cite{LI20121002}, bioinformatics~\cite{10.1093/bib/bbw068}, manufacturing~\cite{s150202774}, and computer vision~\cite{6618951}. The goal of OCC is to train a machine learning (ML) model that distinguishes normal data from anomalous ones. In OCC, instead of having input-output example pairs as in the usual setup for supervised learning, only the input information is provided. Since the training example does not contain the class labels, the OCC is often called semi-supervised learning and is more difficult than the binary or multinomial classification with the label information. Moreover, a multinomial classifier can be constructed with multiple one-class classifiers.

One-class classification problems have been tackled by statistical machine learning approaches, such as principal component analysis (PCA)~\cite{HOFFMANN2007863}, one-class support vector machine (OC-SVM)~\cite{NIPS1999_8725fb77,10.1162/089976601750264965,Tax2004}, and deep learning based algorithms~\cite{chalapathy2019deep,perera2021one}. In particular, an autoencoder, a feed-forward neural network that aims to copy its input to its output~\cite{bourlard_auto-association_1988,690370209,NIPS1993_9e3cfc48}, is widely used in one-class classification. An autoencoder consists of an encoder, which extracts the essential feature of data and reduces dimension, and a decoder, which reconstructs the data. Given a training dataset, an autoencoder is trained to act as an identity function with respect to the training dataset and the mean squared reconstruction error is subject to minimization. For one-class classification, after training a neural network as an autoencoder with normal class data, the reconstruction error can be used as a decision function~\cite{10.1145/2689746.2689747, doi:10.1137/1.9781611974973.11}. Alternatively, the autoencoder can be used as a feature extractor of other statistical machine learning techniques like OC-SVM~\cite{ERFANI2016121, pmlr-v80-ruff18a,chalapathy2018anomaly}.

As a classical autoencoder is able to learn the efficient representation of low dimensional latent space, a quantum autoencoder (QAE) is proposed for efficient quantum data compression. The QAE utilizes a Parameterized Quantum Circuit (PQC)~\cite{Romero_2017}, which is central in variational quantum algorithms~\cite{cerezo2020variational}. In addition to quantum data compression, several applications of QAE have been explored including denoising quantum data~\cite{PhysRevLett.124.130502}, quantum error correction~\cite{locher2022qec}, quantum error mitigation~\cite{PhysRevA.103.L040403}, and quantum metrology~\cite{du2021qae}. The QAE has also been explored for detecting anomalous phases in the context of quantum Hamiltonian problems~\cite{PhysRevResearch.3.043184}.

Motivated by the success of classical autoencoders for OCC problems, we present a Variational Quantum One-Class Classifier (VQOCC) algorithm based on the QAE that applies to classical data. The VQOCC is composed of data encoding, PQC, and quantum measurements for classical post-processing. In the past, anomaly detection algorithms based on the quantum OC-SVM and quantum PCA that could achieve exponential speedup were proposed~\cite{PhysRevA.97.042315}. However, these quantum algorithms require expensive subroutines, such as the quantum linear solver~\cite{HHL} and matrix exponentiation~\cite{qPCA} that are not suitable for Noisy Intermediate-Scale Quantum (NISQ) computing~\cite{Preskill2018quantumcomputingin}. In contrast, training a shallow-depth PQC with a classical optimizer is regarded as a promising approach for near-term quantum machine learning~\cite{RevModPhys.94.015004}. This work focuses on taking the NISQ-friendly approach that constructs a variational quantum algorithm for one-class classification with classical data, and verifying whether a quantum advantage can be attained.

Numerical experiments are performed on handwritten digits and the Fashion-MNIST dataset with open-source Python API Qibo~\cite{qibo2021} for quantum circuit simulation. The performance of VQOCC is evaluated via the area under a receiver operating characteristic (ROC) curve (AUC), and compared to classical methods including OC-SVM, Kernel PCA, and deep convolutional autoencoder (DCAE). We benchmark the performance of VQOCC with various structures of the quantum autoencoder. The structure of the QAE is determined by selecting data encoding, the number of PQC layers, and the size of the latent feature space. The general result of VQOCC shows comparable performance to the classical methods despite having the number of model parameters grow only logarithmically with the data feature size. Notably, the performance of VQOCC is better than DCAE under similar training conditions. 

The remainder of the paper is organized as follows. Section~\ref{sec:oc} describes the one-class classification and reviews some of the well-known approaches to the problem. Section~\ref{sec:qae} explains the quantum autoencoder, which is the basis of the quantum one-class classifier proposed in this work. Section~\ref{sec:VQOCC} explains the application of quantum autoencoder for one-class classification and constructing different models via modifying the ansatz (i.e. structure of the PQC) and cost functions. Numerical experiments performed using scikit-learn and Qibo with handwritten digits and Fashion-MNIST datasets are explained in Sec.~\ref{sec:exp}. This section also compares the AUC of ROC curves of our algorithm with a one-class SVM, a kernel PCA, and a deep convolutional autoencoder. Section~\ref{sec:conclusion} provides conclusion and suggestions for future work.

\section{One-Class Classification}
\label{sec:oc}

\begin{figure*}[ht]
    \centering
    \includegraphics[width=0.95\textwidth]{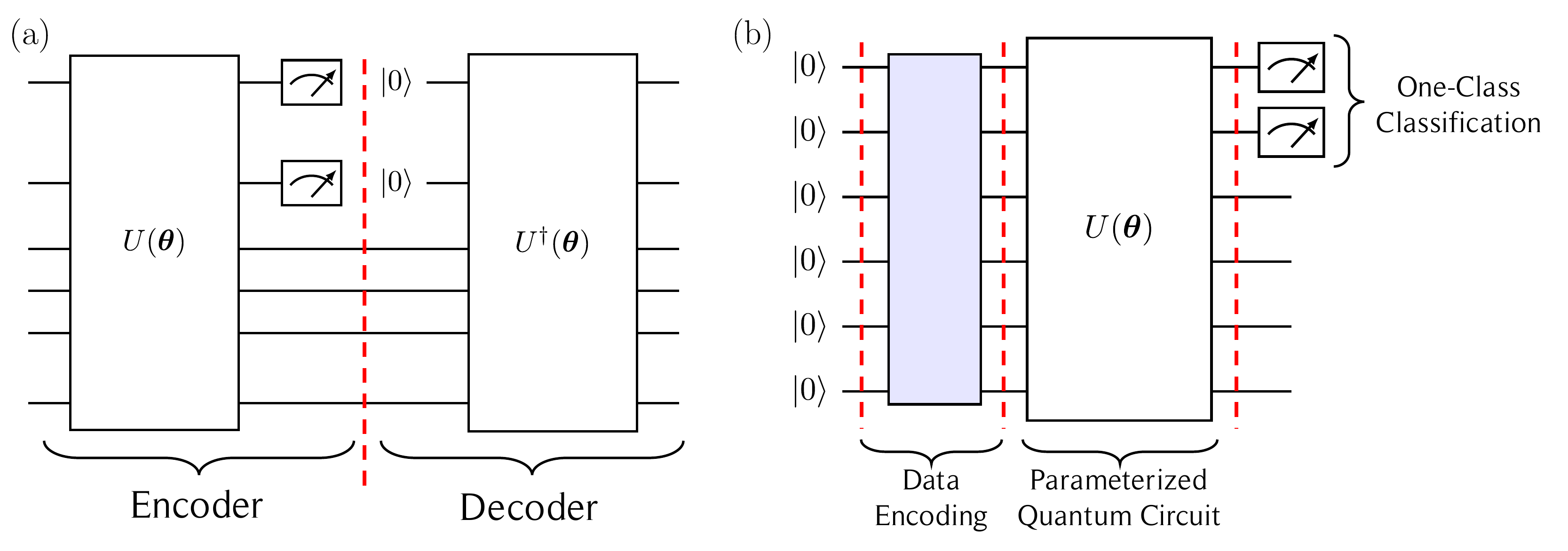}
    \caption{Graphical representation of (a) QAE and (b) VQOCC with number of trash qubits $n_{t}=2$ and total qubits $n=6$. Quantum autoencoder is composed of encoder and decoder parts, which are represented as parameterized quantum circuit $U(\vec{\theta})$ and $U^{\dagger}(\vec{\theta})$, respectively. The VQOCC quantum circuit consists of three parts: data encoding (blue rectangle), parameterized quantum circuit $U(\vec{\theta})$, and measurement of trash qubits for one-class classification. Note that the parameterized quantum circuit and measurement from the VQOCC quantum circuit is directly taken from the encoder part of quantum autoencoder.}
    \label{fig:VQOCC}
\end{figure*}

Assigning an input data to one of a given set of classes is a canonical problem in pattern recognition and can be formally described as a classification problem. Classification aims to predict the class label of an unseen (test) data $\tilde{\vec{x}} \in \mathbb{R}^N$, given a labelled (training) dataset $$\mathcal{D} = \left\{ (\vec{x}_1, y_1), \ldots, (\vec{x}_M, y_M) \right\} \subset \mathbb{R}^N\times\mathbb{Z}_l,$$
where $l$ is the number of classes. The one-class classification is a special case of the aforementioned problem when $l=1$~\cite{MOYA1996463,perera2021one}. In this case, the training dataset is $\mathcal{D}=\{\vec{x}_{1},\vec{x}_{2},\ldots,\vec{x}_{M}\}$, which is treated as a normal class, and the goal is to identify whether a test data $\tilde{\vec{x}}$ is in the normal class or not. Since anomalous data is not used in training, this is known as semi-supervised learning. It is also possible to perform one-class classification with unsupervised methods with an unlabelled dataset under the assumption that most of the test dataset is composed of normal data~\cite{perera2021one,chalapathy2019deep}.

Given a training dataset of normal class $\mathcal{D}$, a decision function $f(\vec{x};\vec{x}_{1},\vec{x}_{2},\ldots,\vec{x}_{M})$ is attained from a one-class classification algorithm, which expresses how far the input data is from the training dataset. If the decision function $f(\tilde{\vec{x}};\vec{x}_{1},\vec{x}_{2},\ldots,\vec{x}_{M})$ has a value smaller than a certain threshold value $C_{th}$ (i.e. $f(\tilde{\vec{x}})<C_{th}$), then $\tilde{\vec{x}}$ is classified as normal. Otherwise, if $f(\tilde{\vec{x}})>C_{th}$, then the test data is classified as anomalous. If $f(\tilde{\vec{x}})=C_{th}$, the decision can be made at random.

Two well-known statistical approaches for addressing one-class classification problems are principal component analysis (PCA)~\cite{HOFFMANN2007863} and support vector machine (SVM)~\cite{NIPS1999_8725fb77,10.1162/089976601750264965,Tax2004}. PCA is a dimensionality reduction technique that projects data $\vec{x}_{i}$ into a lower dimensional subspace such that the projections have the largest variances. The projected space provides reconstructed data $\hat{\vec{x}}_{i}$. The lower dimensional subspace is determined to minimize the reconstruction error $\sum_{i}\|\vec{x}_{i}-\hat{\vec{x}}_{i}\|^{2}$. Once the lower dimensional subspace is chosen, the reconstruction error $f(\vec{x}) = \|\vec{x}-\hat{\vec{x}}\|^{2}$ can be considered as a decision function for one-class classification, since it will be small for normal data and large for anomalous one. The kernel trick can be utilized in PCA to include non-linearity~\cite{10.1162/089976698300017467}.

The support vector machine is a supervised learning model that aims to find a hyperplane that separates two classes of training data with the maximum margin. Thus it is commonly used in binary classification. The SVM can be modified for one-class classification by finding a maximum-margin hyperplane that separates normal data from the origin. This is known as the one-class SVM (OC-SVM)~\cite{NIPS1999_8725fb77,10.1162/089976601750264965}. The decision function of OC-SVM is
\begin{equation}
\label{eq:SVM_dec_func}
    f(\vec{x}) = \langle\vec{w},\Phi(\vec{x})\rangle - b,
\end{equation}
where $\vec{w}$ and $b$ describes the hyperplane and $\Phi$ is the feature map. If the decision function is positive (negative), the corresponding test data is classified as normal (anomalous). 

Alternatively, the SVM can be modified for one-class classification by finding the smallest hypersphere that encapsulates normal data. This is known as the support vector data description (SVDD)~\cite{Tax2004}. After finding the optimal hypersphere, the data located outside of the hypersphere is classified as anomalous. In this case, the decision function can be expressed as
\begin{equation}
\label{eq:SVDD_dec_func}
    f(\vec{x}) =  \|\Phi(\vec{x})-\vec{a}\|^{2} - R,
\end{equation}
where $\vec{a}$ is a center of the hypersphere, and $R$ is a radius of the hypersphere. Note that when the data is normalized to unit norm, the OC-SVM and SVDD become equivalent~\cite{Tax2001}. Intuitively, these methods can be understood as a process of learning the boundary for the normal data and identifying the data outside of the boundary to be anomalies.

\section{Quantum Autoencoder}
\label{sec:qae}

QAE is the quantum-analog of classical autoencoder, for which a PQC learns to reduce the dimensionality of data~\cite{Romero_2017}. The training is carried out through a classical optimization process; hence, it is the classical-quantum hybrid algorithm. The dimensionality reduction means that a quantum autoencoder compresses quantum data into a smaller number of qubits than the input qubits. Following the convention used in classical machine learning, we refer to the set of qubits to which the data is compressed as latent qubits. A QAE is composed of an encoding part and a decoding part as depicted in Fig.~\ref{fig:VQOCC} (a). The encoding part applies a parameterized unitary gate $U(\vec{\theta})$, where $\vec{\theta}$ is a set of trainable parameters, aiming to compress data into latent qubits. Other qubits are discarded after this step (i.e. traced out) and are called trash qubits. The number of trash qubits is denoted by $n_t$. For example, the QAE circuit in Fig.~\ref{fig:VQOCC} (a) uses four latent qubits and two trash qubits. The decoder applies $U^{\dagger}(\vec{\theta})$ on the latent qubits and a reference state $\ket{0}^{\otimes n_t}$ to reconstruct the initial quantum state. For a QAE to be successful, the parameterized unitary gates for encoding should learn to disentangle latent qubits and trash qubits to put them into a product state. This guarantees the reconstruction of the initial quantum state via decoding with a proper ancillary state. The PQC is trained by minimizing a cost function defined with the quantum state fidelity or Hamming distance between the trash qubit system and the target state $\ket{0}^{\otimes n_t}$~\cite{Romero_2017,Cerezo2021,Bravo_Prieto_2021}. More details on the cost function used in this work will be described in the next section.

\section{Variational Quantum One-Class Classifier}
\label{sec:VQOCC}

\begin{figure*}[ht]
    \centering
    \includegraphics[width=0.8\textwidth]{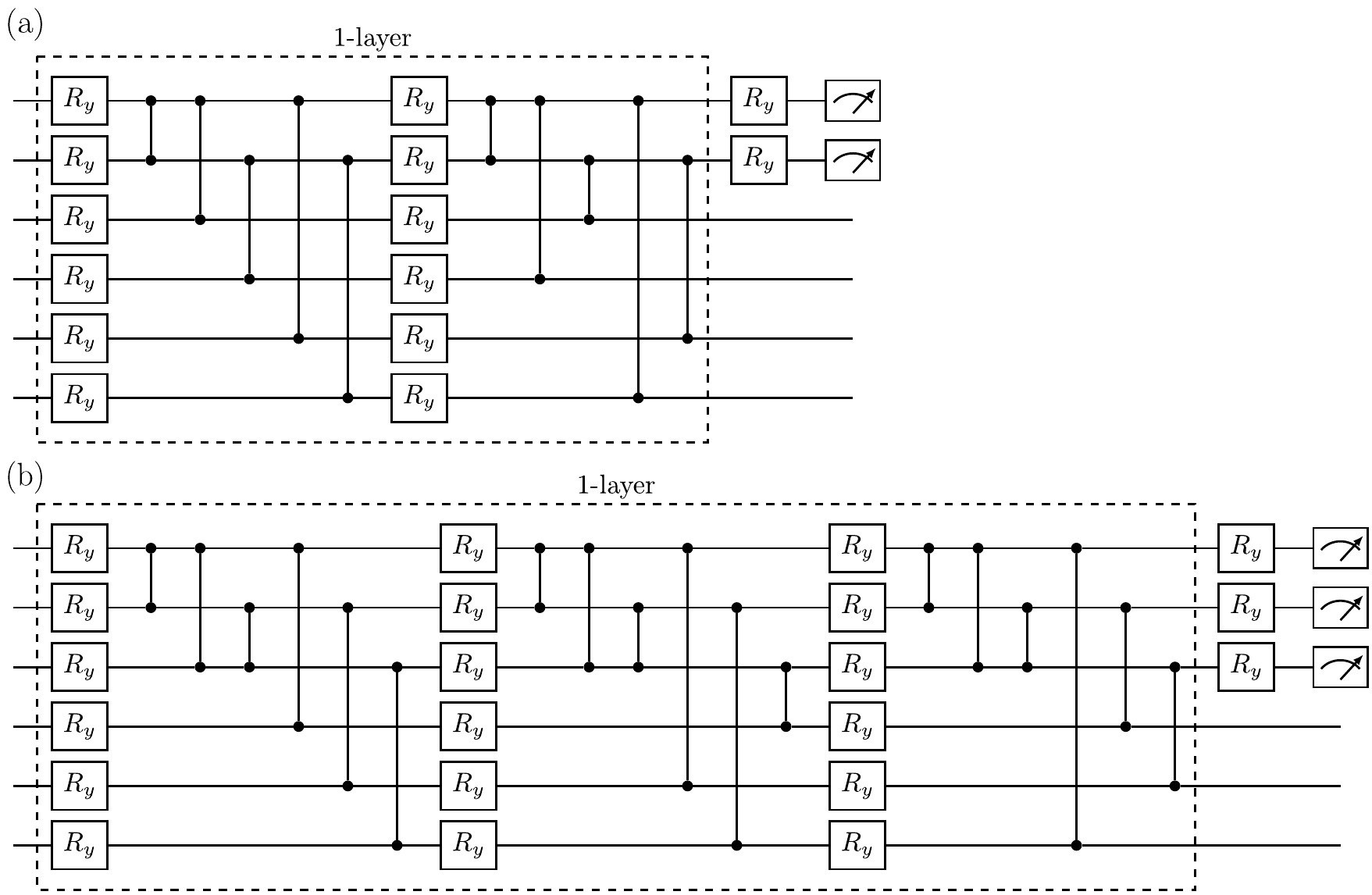}
    \caption{Parameterized quantum circuit ansatz for one layer with number of trash qubits (a) $n_{t}=2$, (b) $n_{t}=3$ and total qubits $n=6$. Dashed boxes correspond to the ansatz of one layer that has $R_y$ rotations and $CZ$ gates between two trash qubits and between a trash qubit and a latent qubit. $CZ$ gates are applied to different combinations of trash-latent qubit pairs after $R_y$ rotations.}
    \label{fig:ansatz}
\end{figure*}

The QAE lays the ground for variational quantum one-class classification. The structure of a QAE can be simplified if it is applied to a one-class classification. Namely, only the encoder part of the QAE is needed. In Romero et al.~\cite{Romero_2017}, two cost functions based on the trash state fidelity and the decoded state fidelity were analyzed. It shows that the trash state fidelity is the upper bound of the decoded state fidelity, and when the trash state fidelity equals one, the decoded state fidelity also equals one. The cost function based on the decoded state requires the access of two identical copies of the input state, whereas the cost function based on the trash state does not. Thus formulating the optimization problem with the cost function that only uses the trash state is more advantageous in terms of computational resources. 

After training the PQC to minimize the cost function for the normal class training dataset, the cost function for anomalous data is expected to yield values far from zero. Hence by setting a threshold value to the cost function, normal and anomalous data can be discriminated. In this case, the cost function can be understood as a decision function $f$ of one-class classification, which is analogous to using the reconstruction error as a decision function in the classical autoencoder.

In the following, the essential steps of the QAE-based VQOCC, namely data encoding, parameterized unitary gates, and measurement of $n_{t}$ qubits from which the cost function is evaluated, are explained in detail. Hereinafter, the structure of quantum gates in a PQC is referred to as ansatz. A pictorial representation of the quantum circuit for VQOCC is shown in Fig. ~\ref{fig:VQOCC} (b).

\subsection{Data encoding}
To handle classical data, a quantum machine learning algorithm must be preceded by a procedure that encodes classical data into quantum states~\cite{PhysRevLett.100.160501,Havlicek2019,PhysRevLett.122.040504,lloyd2020quantum}. The quantum data encoding is a map $\Phi:D\rightarrow W$, where $D\subset\mathbb{R}^{N_c}$ and $W\subset\mathbb{C}^{N_q}$ are the subsets of the real and complex vector spaces in which the classical and quantum data are represented, respectively. The map can be implemented by applying a unitary transformation that is determined by classical data to the initial state $\ket{0}^{\otimes n}$, where $n=\log(N_q)$ is the number of qubits. Typically, $n$ ranges from $\log(N_c)$ to $N_c$ depending on the encoding map $\Phi$~\cite{9259210,araujo_divide-and-conquer_2021,araujo2021configurable}. In this work, two quantum data encoding schemes are used. One is amplitude encoding, which encodes classical data into the amplitudes of quantum states. Another one is called Flexible Representation of Quantum Images (FRQI)~\cite{Le2011}, which encodes classical data into rotation angles. Under these encoding schemes, $n=O(\log(N_c))$. The latter uses one more qubit than the former, but its resource overhead for the classical pre-computation is significantly smaller. More details on these quantum data encoding methods are given in Appendix.~\ref{sec:data_encoding}.

The ansatz of a quantum autoencoder serves as a structure for the sequence of quantum gates that are trained to disentangle trash qubits and latent qubits. The ansatz depicted in Fig.~\ref{fig:ansatz}, which is adapted from Ref.~\cite{Bravo_Prieto_2021}, is constructed to achieve this. It is composed of layers with parameterized single-qubit y-axis rotations $R_y(\theta_j)=e^{-i\theta_j Y/2}$, followed by controlled-$Z$ ($CZ$) gates. The $CZ$ gates are applied between trash-trash qubit pairs and latent-trash qubit pairs. Each layer applies a sequence of $CZ_{\text{latent-trash}}\cdot  CZ_{\text{trash-trash}} \cdot \bigotimes_{j=1}^{n} R_y(\theta_j)$ multiple times within which a different combination of latent-trash qubit pairing is used. In contrast, there are no $CZ$ gates between latent qubits since it does not contribute to disentangling trash qubits from latent qubits. For example, in Fig.~\ref{fig:ansatz} (a), $CZ$ gates are applied between the first latent qubit and either the first or second trash qubit.

In this work, it will be shown that increasing the number of layers can enhance the performance of one-class classification. However, using more layers increases the number of model parameters and gates and the quantum circuit depth. Evaluating these and studying the tradeoff between classification performance and computational resources is of critical importance for the practical application of VQOCC. The number of parameters ($p$) and two-qubit gates ($g_2$), and the circuit depth ($d$) used for VQOCC are 
\begin{equation}
\label{eq:num_params}
    p=n_{t}(nL+1),
\end{equation}
\begin{equation}
    g_2 = \left(\frac{n_{t}^{3}-3n_{t}^{2}}{2}+nn_t\right)L,
\end{equation}
and
\begin{equation}
    d = 1+\left(\frac{n_{t}^{3}-3n_{t}^{2}}{2}+nn_t+n_t\right)L,
\end{equation}
respectively, where $L$ is the number of layers. 

Because $n=O(\log(N_c)$ when amplitude encoding or FRQI encoding is utilized, the number of optimization parameters shown in Eq.~(\ref{eq:num_params}) increases logarithmically with the size of data, which is in stark contrast to a large number of parameters required in classical deep learning algorithms. A potential quantum advantage is rooted in the fact that VQOCC can learn to discriminate anomalous data using only a logarithmic number of model parameters with respect to the number of features that describe the data.

\subsection{Cost function}

The choice of cost function is critical in training PQCs. As described in Sec.~\ref{sec:qae}, minimizing the cost function for a QAE is equivalent to making the trash qubit state as close as possible to the reference state $\ket{0}^{\otimes n_t}$. PQCs are trained such that only normal data can provide the reference state $\ket{0}^{\otimes n_t}$ on trash qubits, whereas anomalous data will provide different states on the trash qubits. In the early developments of QAE~\cite{Romero_2017}, the quantum state fidelity between measured trash qubit state and $\ket{0}^{\otimes n_{t}}$ was chosen as a cost function. However, Ref. ~\cite{Cerezo2021} showed that cost functions with global observables induce exponentially vanishing gradients, so called barren plateaus~\cite{McClean2018}, even in shallow quantum circuits. Consequently, the fidelity-based cost function used in the initial work is subject to the unwanted barren plateau effect. One way to avoid this is to construct a cost function with local observables~\cite{Cerezo2021}. The localized cost function used in this paper is based on the Hamming distance between the measurement outcome in the computational basis and the reference state $\ket{0}^{\otimes n_{t}}$~\cite{Bravo_Prieto_2021}. More specifically, the local cost function based on the Hamming distance can be written as
\begin{equation}
\label{eq:cost}
    C  = \frac{1}{2}\sum_{j=1}^{n_{t}}(1-\langle Z_j\rangle),
\end{equation}
where $\langle Z_{j}\rangle$ is an expectation value of the Pauli-$Z$ operator for the $j$th trash qubit. This cost function is zero when the trash qubit state is disentangled from latent qubits and equals to the reference state $\ket{0}^{\otimes n_{t}}$, which indicates the compression of the quantum state to latent qubits.

While keeping the local property, a cost function can be constructed in a different manner. For instance, a cost function can be formulated in terms of a log loss function as
\begin{equation}
\label{eq:cost_log}
    C =\sum_{j=1}^{n_{t}} \log \left( \frac{1-\langle Z_j\rangle}{2}\right).
\end{equation}

\section{Numerical experiments}
\label{sec:exp}

\begin{figure*}[ht]
    \centering
    \includegraphics[width=0.9\textwidth]{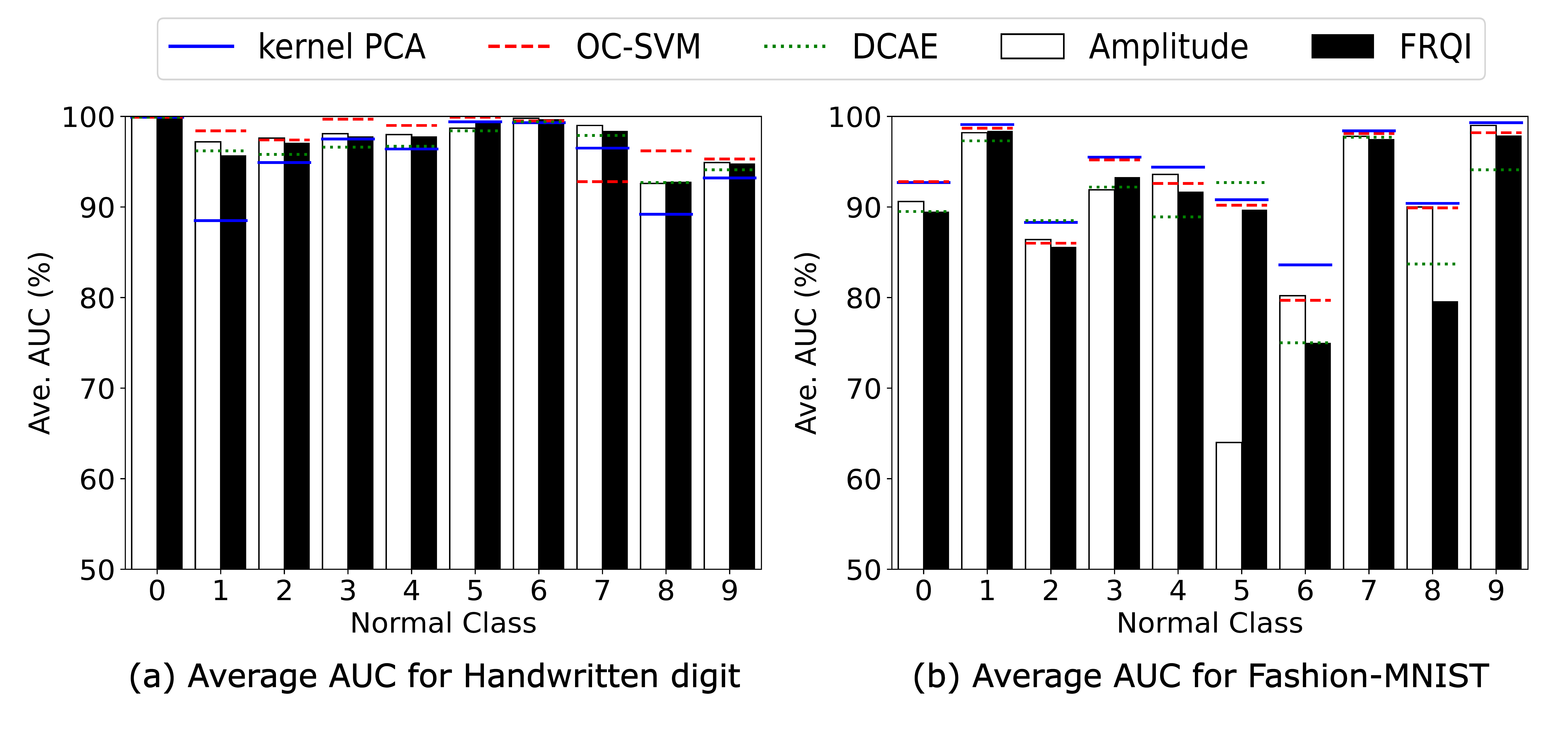}
    \caption{Average AUCs (over 10 seeds) of VQOCC on handwritten digit and Fashion-MNIST dataset for ten different normal classes. We report the best results from different numbers of trash qubits $n_{t}$ and layers $L$ of parameterized quantum circuits with two quantum data encoding schemes, amplitude encoding (open) and FRQI (filled). The results of kernel PCA (blue solid), OC-SVM (red dashed), and DCAE (green dotted) are illustrated as baselines.
    }
    \label{fig:best_auc}
\end{figure*}

\begin{table*}[ht]
\centering
\begin{tabular}{cccccccccc}
\hline
\multicolumn{2}{c}{}                                                                                                                                    & \multicolumn{8}{c}{AUC(\%)}                                                                                                                                                                                                                                                  \\ \hline
\multirow{2}{*}{Dataset}                                                      & \multirow{2}{*}{\begin{tabular}[c]{@{}c@{}}Normal\\ Class\end{tabular}} & \multicolumn{2}{c}{Amplitude}            & \multicolumn{2}{c}{FRQI}                 & \multicolumn{2}{c}{DCAE}                  & \multirow{2}{*}{\begin{tabular}[c]{@{}c@{}}Kernel\\ PCA\end{tabular}} & \multirow{2}{*}{\begin{tabular}[c]{@{}c@{}}OC-\\ SVM\end{tabular}} \\ \cline{3-8}
                                                                              &                                                                         & Average       & \multicolumn{1}{c}{Best} & Average       & \multicolumn{1}{c}{Best} & Average        & \multicolumn{1}{c}{Best} &                                                                       &                                                                    \\ \hline
\multirow{10}{*}{\begin{tabular}[c]{@{}c@{}}Handwritten\\ Digit\end{tabular}} & \multicolumn{1}{c|}{0}                                                  & $99.9\pm0.1$  & 100.0                    & $99.9\pm0.1$  & 100.0                    & $99.9 \pm 0.1$ & 100.0                    & $99.9$                                                                & $99.9$                                                             \\
                                                                              & \multicolumn{1}{c|}{1}                                                  & $97.2\pm 0.8$ & 98.7                     & $95.7\pm 0.9$ & 97.0                     & $96.2 \pm 1.2$ & 97.9                     & $88.5$                                                                & $98.4$                                                             \\
                                                                              & \multicolumn{1}{c|}{2}                                                  & $97.6\pm 0.9$ & 99.0                     & $97.1\pm 0.9$ & 98.7                     & $95.8 \pm 2.3$ & 99.4                     & $94.9$                                                                & $97.4$                                                             \\
                                                                              & \multicolumn{1}{c|}{3}                                                  & $98.1\pm 0.3$ & 98.7                     & $97.8\pm 0.5$ & 98.4                     & $96.6 \pm 1.1$ & 99.5                     & $97.5$                                                                & $99.7$                                                             \\
                                                                              & \multicolumn{1}{c|}{4}                                                  & $98.0\pm 0.6$ & 99.2                     & $97.8\pm 0.2$ & 98.1                     & $96.7 \pm 1.1$ & 99.0                     & $96.4$                                                                & $99.0$                                                             \\
                                                                              & \multicolumn{1}{c|}{5}                                                  & $98.7\pm 0.5$ & 99.5                     & $99.3\pm 0.3$ & 99.6                     & $98.4 \pm 1.1$ & 99.8                     & $99.4$                                                                & $99.9$                                                             \\
                                                                              & \multicolumn{1}{c|}{6}                                                  & $99.8\pm 0.1$ & 100.0                    & $99.7\pm 0.1$ & 99.9                     & $99.4 \pm 0.3$ & 100.0                    & $99.3$                                                                & $99.5$                                                             \\
                                                                              & \multicolumn{1}{c|}{7}                                                  & $99.0\pm 0.5$ & 99.7                     & $98.4\pm 0.9$ & 99.3                     & $97.9 \pm 1.0$ & 99.4                     & $96.5$                                                                & $92.8$                                                             \\
                                                                              & \multicolumn{1}{c|}{8}                                                  & $92.6\pm 1.8$ & 95.2                     & $92.8\pm 2.2$ & 96.0                     & $92.7 \pm 2.1$ & 96.7                     & $89.2$                                                                & $96.2$                                                             \\
                                                                              & \multicolumn{1}{c|}{9}                                                  & $94.9\pm 1.3$ & 96.5                     & $94.8\pm 0.5$ & 95.2                     & $94.1 \pm 1.5$ & 96.6                     & $93.2$                                                                & $95.3$                                                             \\ \hline
\multirow{10}{*}{\begin{tabular}[c]{@{}c@{}}Fashion-\\ MNIST\end{tabular}}    & \multicolumn{1}{c|}{0}                                                  & $90.6\pm 0.6$ & 91.4                     & $89.5\pm 1.1$ & 90.4                     & $89.5 \pm 4.1$ & 96.1                     & $92.7$                                                                & $92.8$                                                             \\
                                                                              & \multicolumn{1}{c|}{1}                                                  & $98.2\pm 0.2$ & 98.7                     & $98.4\pm 0.1$ & 98.8                     & $97.3 \pm 1.1$ & 99.4                     & $99.1$                                                                & $98.7$                                                             \\
                                                                              & \multicolumn{1}{c|}{2}                                                  & $86.4\pm 0.6$ & 87.6                     & $85.6\pm 0.6$ & 87.3                     & $88.5 \pm 3.2$ & 95.7                     & $88.3$                                                                & $86.0$                                                             \\
                                                                              & \multicolumn{1}{c|}{3}                                                  & $91.9\pm 1.0$ & 93.7                     & $93.3\pm 0.6$ & 93.9                     & $92.2\pm1.3$   & 95.0                     & $95.5$                                                                & $95.2$                                                             \\
                                                                              & \multicolumn{1}{c|}{4}                                                  & $93.6\pm 0.2$ & 93.9                     & $91.7\pm 1.4$ & 92.7                     & $88.9\pm4.2$   & 97.6                     & $94.4$                                                                & $92.6$                                                             \\
                                                                              & \multicolumn{1}{c|}{5}                                                  & $64.0\pm 2.5$ & 70.1                     & $89.7\pm 0.5$ & 90.4                     & $92.7 \pm 1.4$ & 94.4                     & $90.8$                                                                & $90.2$                                                             \\
                                                                              & \multicolumn{1}{c|}{6}                                                  & $80.2\pm 0.4$ & 80.8                     & $75.0\pm 2.0$ & 77.5                     & $75.0 \pm 3.7$ & 96.3                     & $83.6$                                                                & $79.7$                                                             \\
                                                                              & \multicolumn{1}{c|}{7}                                                  & $97.8\pm 0.3$ & 98.2                     & $97.6\pm 0.1$ & 97.9                     & $97.5 \pm 0.4$ & 98.8                     & $98.4$                                                                & $98.1$                                                             \\
                                                                              & \multicolumn{1}{c|}{8}                                                  & $90.0\pm 1.0$ & 91.4                     & $79.6\pm 3.3$ & 84.0                     & $83.7 \pm 3.9$ & 94.5                     & $90.4$                                                                & $89.9$                                                             \\
                                                                              & \multicolumn{1}{c|}{9}                                                  & $99.0\pm 0.3$ & 99.4                     & $97.9\pm 0.6$ & 98.6                     & $94.1\pm4.5$   & 98.9                     & $99.3$                                                                & $98.2$                                                            
\end{tabular}
\caption{The average and best AUCs in \% and one standard deviation (over 10 seeds) of the one-class classification on handwritten digit and Fashion-MNIST dataset with variational quantum one-class classifier (VQOCC) with both amplitude and FRQI encoding, and deep convolutional autoencoder (DCAE). For VQOCC results, the better results are reported while varying the number of trash qubits $n_{t}$ and layers of parameterized quantum circuits $L$. The results from Kernel PCA and OC-SVM are also shown as a competing method. }
    \label{tab:AUC}
\end{table*}

\begin{figure*}[ht]
    \centering
    \includegraphics[width=1.0\textwidth]{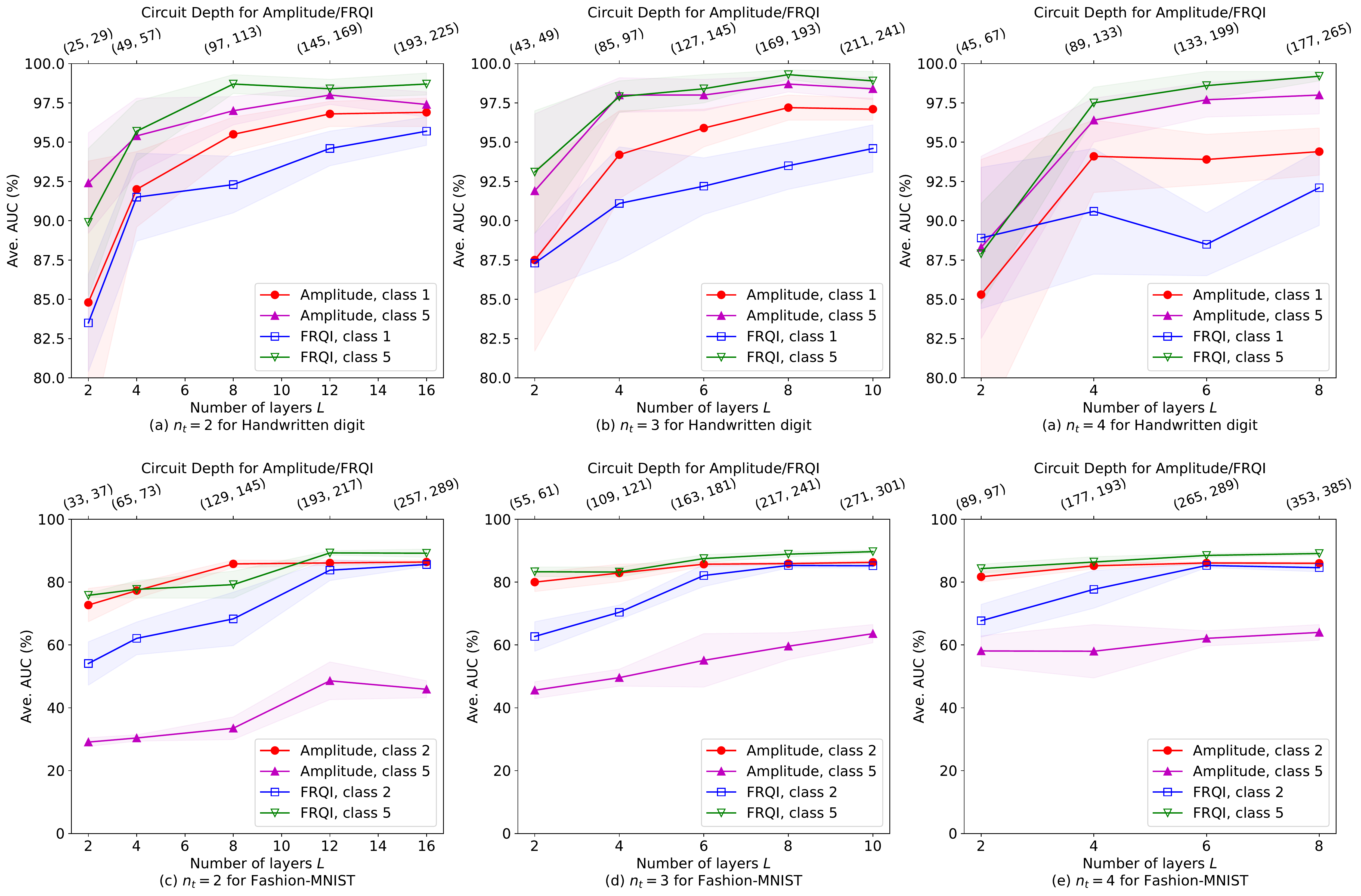}
    \caption{Average AUCs and their standard deviations of VQOCC varying the number of trash qubits $n_{t}=2,3,4$ and layers $L$ on handwritten digits (a, b, c) and Fashion-MNIST (d, e, f) dataset. Shaded areas indicate the standard deviations. For handwritten digits, the results from class 1 and 5 are shown, and for Fashion-MNIST, the results form class 2 and 5 are shown as representative examples. The plots include results for two different encoding schemes, amplitude encoding and FRQI encoding. The circuit depths at a number of layers $L$ for different encoding schemes are indicated on the top of each plot.}
    \label{fig:nLdependencies}
\end{figure*}

This section reports the benchmarking results for VQOCC, obtained by classical simulations carried out with open source framework Qibo~\cite{qibo2021}. The simulation is performed with two datasets, an $8\times8$ handwritten digits dataset available in scikit-learn~\cite{JMLR:v12:pedregosa11a}, and the Fashion-MNIST dataset~\cite{xiao2017fashionmnist} downsampled to $16\times16$ pixels. Both datasets have ten different classes. Samples of the datasets are depicted in Appendix~\ref{sec:data_samples}. In each numerical simulation, one class is treated as the normal class, and the training sample size is 100. After training, both normal and anomalous data samples are used as a test dataset. An equal number of samples from each class is used as test samples. A test dataset is composed of 70 and 100 test samples from each class, and consequently, the total number of test datasets used in the experiment is 700 and 1000 for handwritten digits and Fashion-MNIST datasets, respectively.

Besides the datasets, there are three variables in the numerical experiments: quantum data encoding methods, the number of trash qubits denoted by $n_{t}$, and the number of PQC layers denoted by $L$. Amplitude encoding and FRQI encoding are used for quantum data encoding, as described before. For amplitude encoding, 6-qubit states and 8-qubit states are used for handwritten digits and Fashion-MNIST datasets, respectively. For FRQI encoding, 7-qubit states and 9-qubit states are used for handwritten digits and Fashion-MNIST datasets, respectively. Three different numbers of trash qubits, $n_{t}=2,3,4$, are used in the simulation. For each $n_{t}$, different values of $L$ are used. More specifically, $L\in\lbrace 2, 4, 8, 12, 16\rbrace$ for $n_{t}=2$, $L\in\lbrace 2, 4, 6, 8, 10\rbrace$ for $n_{t}=3$, and $L\in\lbrace 2, 4, 6, 8\rbrace$ for $n_{t}=4$. The maximum $L$ for each $n_{t}$ is chosen in a way that the maximum circuit depths for all $n_{t}$ are approximately the same for a fair comparison among different $n_{t}$. The evaluation metric used for numerical experiments is the Area Under the Receiver Operating Characteristic (ROC) curve (AUC), which is commonly used for one-class classifiers~\cite{perera2021one}.

The optimization of the parameterized quantum circuit was performed with the Adam optimizer~\cite{kingma2017adam} in Tensorflow~\cite{tensorflow2015-whitepaper}. Tensorflow was set to be the simulation backend of Qibo, which enables the automatic differentiation for the computation of gradients. We used mini-batch gradient descent with a batch size of 10. The learning rate was set to be 0.1, and the number of iterations was 150. 
%We selected the $n_t$ and $L$ that showed better results for each case and reported the better results. 

The results presented in this section are based on minimizing the Hamming distance cost function shown in Eq.~(\ref{eq:cost}). Experiments based on the logarithmic cost function shown in Eq.~(\ref{eq:cost_log}) tested on the handwritten digit and the Fashion-MNIST datasets with amplitude encoding produced similar results that are reported in Appendix \ref{sec:log_cost}. 

We compared the performance of VQOCC with various classical methods, such as the Kernel PCA, OC-SVM, and deep neural network. For Kernel PCA and OC-SVM, the Gaussian radial basis function (RBF) kernel was used. We selected the inverse length parameter $\gamma$ from $\gamma \in \{ 2^{-10},2^{-9},\ldots,2^{-1}\}$ via grid search using the performance on a small holdout set (10$\%$ of randomly drawn normal and anomalous test samples), and for OC-SVM, we chose the hyperparameter $\nu \in \{0.01,0.1\}$, which represents the lower (upper) bound for the fraction of support vectors (errors), and report the better result.
 
As a deep learning method, a deep convolutional autoencoder (DCAE)~\cite{Masci11stackedconvolutional,pmlr-v80-ruff18a} was chosen as a baseline. The size of the deep architecture was constrained so that the number of model parameters is comparable to that of VQOCC for a fair comparison. A detailed description of the DCAE architecture is described in Appendix~\ref{sec:dcae}. The number of parameters used in DCAE is 158 and 254 for the handwritten digit and the Fashion-MNIST datasets, respectively.

\subsection{Results overview}

The main results are presented in Fig.~\ref{fig:best_auc} and Table~\ref{tab:AUC}. Each value in the figure and table represents the result obtained with the best $n_{t}$ and $L$. In Fig.~\ref{fig:best_auc}, two data encoding methods, amplitude encoding and FRQI encoding, are shown as open and filled bars, respectively.  The results obtained from Kernel PCA and OC-SVM, and DCAE are displayed as blue solid, red dashed, and green dotted lines, respectively. In Table~\ref{tab:AUC}, the average and best AUCs of the one-class classification on handwritten digits and Fashion-MNIST datasets are provided. In general, the results from both encoding methods are on par with the classical algorithms and show better performances than DCAE. The cases from the class 2, 6, and 7 of the handwritten digit dataset even show that the performances from the quantum algorithm outperform the performances from all three classical algorithms. Also note that the best AUCs from VQOCC with either of the encoding methods at the class 0, 1, 2, 4, 6, 7, and 9 of handwritten digit and the class 8 and 9 of Fashion-MNIST show better performances than the AUCs from both OC-SVM and Kernel PCA. The best AUCs from DCAE are generally better than the best AUCs from VQOCC in the Fashion-MNIST dataset, but the standard deviations of DCAE are larger than that of VQOCC. The two quantum data encoding methods performed comparably, except one case at the class 5 of Fashion-MNIST with the significantly lower performance of 64.0$\%$, due to the data normalization in amplitude encoding (Appendix.~\ref{sec:nor_data}). Note that even though FRQI encoding requires one more qubit than amplitude encoding, quantum state preparation of image data is simpler with FRQI encoding (see Appendix.~\ref{sec:data_encoding}). 

\subsection{Effects of the QAE structure}

The performances of VQOCC are evaluated for various QAE structures determined by different values of $n_{t}$ and $L$. Figure~\ref{fig:nLdependencies} illustrates the average AUCs of VQOCC for various $n_{t}$ and $L$ with standard deviations as colored shade on the handwritten digit and Fashion-MNIST datasets. We chose a few classes to illustrate the dependencies. Class 1 and 5 of handwritten digits and class 2 and 5 of Fashion-MNIST were chosen with both encoding schemes. The circuit depths at each $L$ are also shown at the top of each plot. The general trend shows that the average AUC (i.e. the classification performance) increases with $L$. Furthermore, not only does the performance increase but also the standard deviation of AUCs decreases as $L$ increases. This implies that the PQC with low depth is more likely to fall into a local minimum, which impedes the classification performance. 

The results in Figure~\ref{fig:nLdependencies} show that the performance saturates as $L$ increases. In general, an average AUC increases with $L$, but in some cases, the performance hits a plateau when $L$ is beyond a certain threshold $L_{th}$. For instance, $L_{th}$ of class 5 of the handwritten digit dataset is $L_{th}=8$ and $L_{th}=4$ for $n_{t}=2$ and $3$, respectively, for both encoding schemes, whereas in the case of $n_{t}=4$, the performance plateau was not observed in class 5 of the handwritten digit dataset. Such behaviors are also observed at the class 1 of the handwritten digit dataset for Amplitude encoding at $n_{t}=4$, the class 5 of Fashion-MNIST dataset for both encoding schemes at $n_{t}=2$, and the class 2 of Fashion-MNIST dataset for both encoding schemes at $n_{t}=2,3,$ and $4$. Recall that the number of model parameters increases linearly with $L$. This observation shows the relevance of the over-parameterization in variational quantum machine learning~\cite{larocca2021theory}.

\section{Conclusion}
\label{sec:conclusion}
We proposed a semi-supervised variational QML algorithm for one-class classification of classical data. In general, one-class classification problems are considered to be more difficult than traditional binary or multi-class classification problems due to the absence of labels in the dataset. Consequently, it requires a different approach. Our algorithm, dubbed VQOCC, utilizes a fully-parameterized quantum autoencoder (QAE) that learns to extract essential features of normal data. Unlike the conventional QAE, VQOCC only uses the encoding part, which is trained to recognize anomalous data by minimizing a loss function defined with trash qubits left out from the data compression. Since the algorithm is based on training a fully parameterized quantum circuit and only uses half the circuit of the usual QAE, it is an excellent candidate for the NISQ application.

We explored the performance of the VQOCC algorithm for the handwritten digits and Fashion-MNIST datasets by varying the classifier structure, which includes the data encoding scheme, the number of layers in the ansatz, and the number of measured trash qubits. After tailoring the structure of the VQOCC algorithm, the performance was compared to classical one-class classification methods including OC-SVM, Kernel PCA, and deep convolutional autoencoder (DCAE). The number of parameters in DCAE was matched to that in PQCs for a fair comparison. VQOCC generally performed better than DCAE and was comparable to OC-SVM and Kernel PCA. However, it is unsure whether increasing the number of parameters in PQCs by having more layers will constantly improve the performance to significantly surpass classical methods. Numerical experiments show that increasing the number of circuit layers (and hence the number of model parameters) is deemed to be effective in enhancing performance until it hits the plateau when a certain value is reached.

Note that statistical methods like OC-SVM and Kernel PCA are known to be less effective at complex high-dimensional datasets ~\cite{Bengio07scalinglearning} and deep learning algorithms are shown to be very effective at this problem. VQOCC proposed in this work can be an alternative model of deep learning algorithms and its classification capabilities for complex high-dimensional datasets are left for future investigation. Since the number of qubits for encoding classical data can be exponentially smaller than the number of features, the model parameters in VQOCC can be exponentially smaller than classical deep learning algorithms. Moreover, OC-SVM and Kernel PCA require at least $O(d)$ computational complexity for a $d$-dimensional dataset, in stark contrast to $O(\log d)$ number of qubits and model parameters required in VQOCC.

Interesting future work is to apply different ansatz of PQCs for VQOCC. One particular example of ansatz is quantum circuits with a hierarchical structure~\cite{grant_hierarchical_2018}. Previously, the hierarchical structure was successful for binary classification~\cite{grant_hierarchical_2018,hur2021quantum}, and hence it is natural to extend it to one-class classification. Another promising direction could be a combination of a QAE and OC-SVM, possibly with quantum kernels. A similar approach exists in classical machine learning in which an autoencoder is used as a feature extractor~\cite{ERFANI2016121} and OC-SVM is used for anomaly detection on the compressed data. Appendix~\ref{sec:nor_data} shows that the classification results of OC-SVM and Kernel PCA are affected under the data normalization, which is required in amplitude encoding. This implies that the VQOCC can be further improved by carefully choosing the data encoding scheme which must be preceded by classical data preprocessing. Finally, it is worth mentioning that our approach can be extended to the unsupervised learning of one-class classification. In the classical unsupervised setup, most of the data is assumed to be comprised of normal data and anomalous data will show a high false positive rate~\cite{chalapathy2019deep, pmlr-v80-ruff18a}. This assumption can be directly applied to our VQOCC algorithm. 

\section*{Acknowledgments}
This research was supported by the Yonsei University Research Fund of 2022 (2022-22-0124), by the National Research Foundation of Korea (Grant Nos. 2019R1I1A1A01050161, 2021M3H3A1038085, 2019M3E4A1079666, and 2022M3E4A1074591), and by the KIST Institutional Program (2E31531-22-076).

% %Bibliography
% \bibliographystyle{unsrt}  
% \bibliography{references}  

%
%
%

\appendix
\section{Data encoding}
\label{sec:data_encoding}

\subsection{Amplitude encoding}
\label{sec:AE}
Amplitude encoding loads classical data into the amplitudes of a quantum state. To encode a $2^{n}\times 2^{n}$ pixel image into a quantum state, the input can be represented as $x=(x_{1},x_{2},\ldots,x_{N})^{\top}$, where $N=2^{2n}$, and it can be encoded in a $2n$-qubit quantum state $\phi(x)$ as follows, 

\begin{equation}
\label{eq:amplitude}
    \ket{\phi(x)} = \frac{1}{\|x\|} \sum_{i=1}^{N} x_i \ket{i}.
\end{equation}

\subsection{FRQI : Flexible Representation of Quantum Images}

The flexible representation of quantum images(FRQI)~\cite{Le2011} is designed to encode classical image data into quantum states. Unlike amplitude encoding, which requires nonintuitive routine to prepare arbitrary amplitudes, FRQI encoding is composed of Hadamard and controlled rotation with rotation angles simply achieved from input image. When input is given by $x=(x_{1},x_{2},\ldots,x_{N})^{\top}$, each pixels can be normalized to be in the range of $\theta_{i} \in [0,\frac{\pi}{2}]$, or $\theta=(\theta_{1},\theta_{2},\ldots,\theta_{N})^{\top}$. These angles can be encoded in $(2n+1)$-qubit quantum state $I(\theta)$ as follows,

\label{sec:FRQI}
\begin{equation}
\label{eq:FRQI}
   \ket{I(\theta)} = \frac{1}{2^{n}} \sum_{i=0}^{2^{2n}-1} (\cos( \theta_{i})\ket{0}+\sin(\theta_{i})\ket{1} ) \otimes \ket{i}.
\end{equation}
 
It requires one more qubit to encode classical data, but it is more intuitive and flexible on encoding classical image data into quantum states.

\section{Logarithmic cost function}
\label{sec:log_cost}
In Sec. \ref{sec:VQOCC} of the main text, the local cost function was introduced that it can avoid barren plateaus. The cost function mainly used in the main text is the Hamming distance based cost function \eqref{eq:cost}, but the local cost function can also be constructed in therms of logarithmic function, as \eqref{eq:cost_log}. We here report the AUC result for the handwritten digit and Fashion-MNIST dataset with amplitude encoding and logarithmic cost function in Tab. \ref{tab:log_cost}. The result for logarithmic cost does not show a significant difference to the result from Hamming distance based cost function.

\begin{table}[ht]
\begin{tabular}{ccccc}
\hline
\multicolumn{2}{c}{} & \multicolumn{3}{c}{Average AUC(\%)}\\ 
\hline
Dataset                                                                       & \begin{tabular}[c]{@{}c@{}}Normal\\ Class\end{tabular} & Amplitude & \begin{tabular}[c]{@{}c@{}}Kernel\\ PCA\end{tabular} & \begin{tabular}[c]{@{}c@{}}OC-\\ SVM\end{tabular} \\ \hline
\multirow{10}{*}{\begin{tabular}[c]{@{}c@{}}Handwritten\\ Digit\end{tabular}} & \multicolumn{1}{c|}{0}                                 & $99.9  \pm 0.1$                                            & 99.9                                                 & 99.9                                              \\
                                                                              & \multicolumn{1}{c|}{1}                                 & $97.7 \pm 0.7$                                              & 88.5                                                 & 98.4                                              \\
                                                                              & \multicolumn{1}{c|}{2}                                 & $96.5  \pm 1.2$                                             & 94.9                                                 & 97.4                                              \\
                                                                              & \multicolumn{1}{c|}{3}                                 & $98.0   \pm 0.8$                                            & 97.5                                                 & 99.7                                              \\
                                                                              & \multicolumn{1}{c|}{4}                                 & $97.7   \pm 0.3$                                            & 96.4                                                 & 99.0                                              \\
                                                                              & \multicolumn{1}{c|}{5}                                 & $98.8 \pm 0.6$                                             & 99.4                                                 & 99.9                                              \\
                                                                              & \multicolumn{1}{c|}{6}                                 & $99.8  \pm 0.1$                                            & 99.3                                                 & 99.5                                              \\
                                                                              & \multicolumn{1}{c|}{7}                                 & $98.9   \pm 0.5$                                           & 96.5                                                 & 92.8                                              \\
                                                                              & \multicolumn{1}{c|}{8}                                 & $92.5  \pm 1.3$                                            & 89.2                                                 & 96.2                                              \\
                                                                              & \multicolumn{1}{c|}{9}                                 & $93.7  \pm 2.2$                                            & 93.2                                                 & 95.3    \\
\hline                                                                              
\multirow{10}{*}{\begin{tabular}[c]{@{}c@{}}Fashion-\\ MNIST\end{tabular}} & \multicolumn{1}{c|}{0}                                 & $90.9 \pm 0.3$      & 89.0                                                 & 87.2                                              \\
                                                                           & \multicolumn{1}{c|}{1}                                 & $98.2\pm 0.1$      & 98.5                                                 & 91.3                                              \\
                                                                           & \multicolumn{1}{c|}{2}                                 & $86.8\pm0.4$      & 86.2                                                 & 87.8                                              \\
                                                                           & \multicolumn{1}{c|}{3}                                 & $92.4\pm0.5$      & 91.0                                                 & 91.9                                              \\
                                                                           & \multicolumn{1}{c|}{4}                                 & $92.9\pm0.3$      & 93.6                                                 & 93.6                                              \\
                                                                           & \multicolumn{1}{c|}{5}                                 & $65.0\pm3.5$      & 75.3                                                 & 43.2                                              \\
                                                                           & \multicolumn{1}{c|}{6}                                 & $79.9\pm0.5$      & 79.9                                                 & 80.4                                              \\
                                                                           & \multicolumn{1}{c|}{7}                                 & $97.9\pm0.2$      & 98.1                                                 & 97.0                                              \\
                                                                           & \multicolumn{1}{c|}{8}                                 & $89.8\pm2.0$      & 89.4                                                 & 80.8                                              \\            & \multicolumn{1}{c|}{9}                                 & $99.0\pm0.3$      & 98.9                                                 & 98.3                                                          
\end{tabular}
\caption{Average AUCs in \% and one standard deviation (over 10 seeds) of the one-class classification on the handwritten digit dataset with logarithmic cost function. }
\label{tab:log_cost}
\end{table}

\section{Data samples}
\label{sec:data_samples}

\begin{figure*}[ht]
    \centering
    \includegraphics[width=0.96\textwidth]{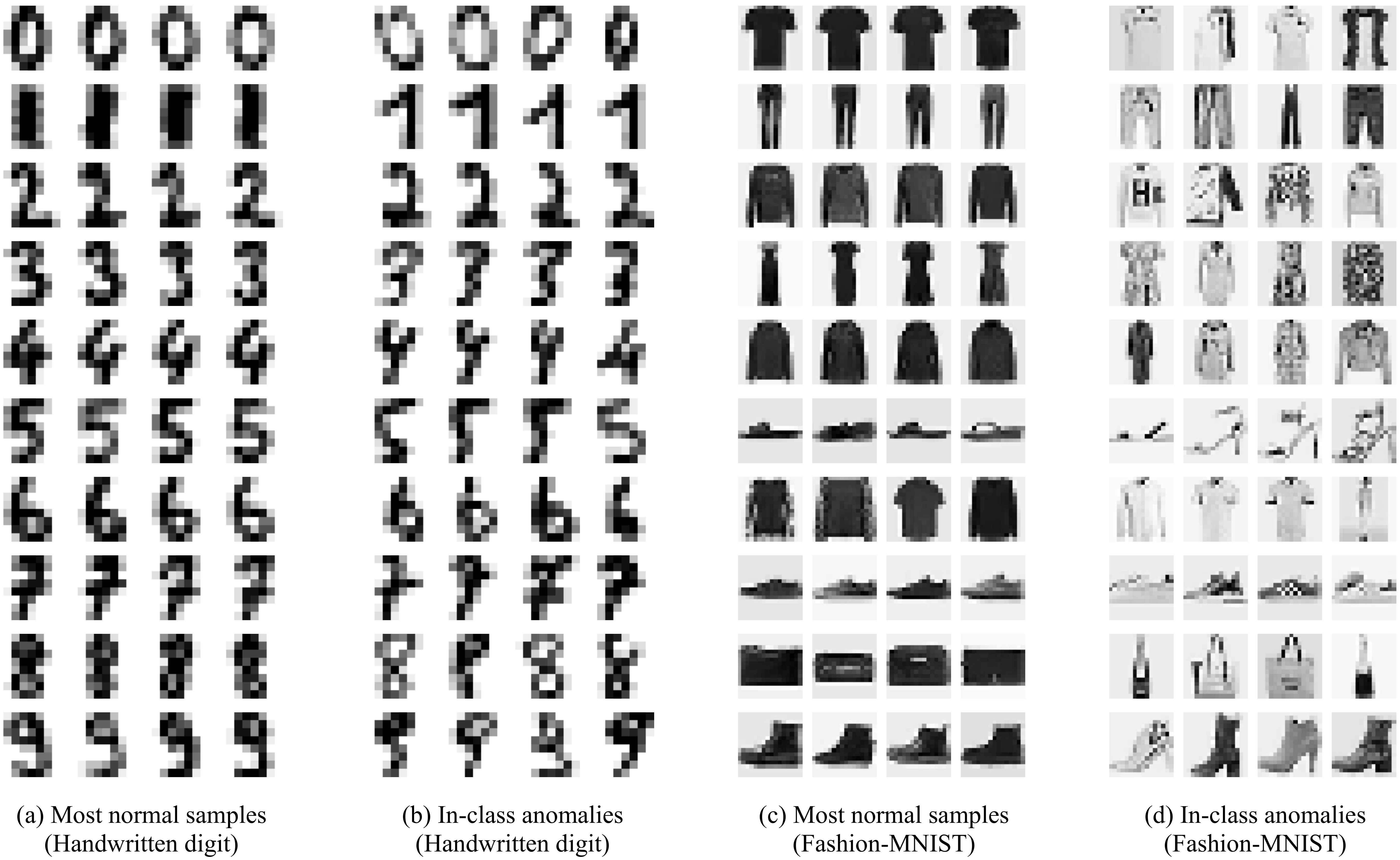}
    \caption{Most normal (a, c) and most anomalous in-class (b, d) samples in the handwritten digit and Fashion-MNIST dataset, respectively, classified by Variational Quantum One-Class Classifier (VQOCC).}
    \label{fig:dataset_display}
\end{figure*}

Figure~\ref{fig:dataset_display} (a), (b), (c), and (d) show the most normal and the most anomalous in-class samples in the handwritten digit data, and the most normal and the most anomalous in-class samples in the Fashion-MNIST data, respectively, determined by the cost function of the VQOCC algorithm.

\section{Deep convolutional autoencoder}
\label{sec:dcae}

The deep learning architecture for DCAE is determined to match the number of parameter in deep learning model to be comparable to the number of parameter used in VQOCC quantum circuit model for fair comparison. The DCAE model is composed of encoder and decoder. For the encoder part of DCAE, we use LeNet-type convolutional neural networks(CNNs), where each convolutional layers are followed by leaky ReLu activation functions and $2\times2$ max-pooling. The decoder has a symmetrical structure to the encoder, where the max-pooling is substituted by upsampling. The encoder architecture is consisted of $2 \times (3\times3\times1)$ filters followed by $2 \times (3\times3\times1)$ filters, and a final dense layer of 4 units. For optimization, the Adam optimizer is used and the Batch Normalization is applied as ~\cite{pmlr-v80-ruff18a}. We train the model with the learning rate $\eta$ from $\eta \in \{1.0 \times10^{-4}, 5.0\times10^{-4}, 1.0 \times10^{-3}, 5.0\times10^{-3}\}$, the weight decay hyperparameter $\lambda$ from $\lambda \in \{10^{-6},10^{-5}\}$, batch size of 10, and 250 epochs, and report the better results.

\section{Effect of data normalization}
\label{sec:nor_data}

Amplitude encoding requires data to be normalized. In this section, we compare the VQOCC result from amplitude encoding to the result from Kernel PCA and OC-SVM with the normalized input data in Tab.~\ref{tab:norm_data}. A few cases show a decrease in performance after normalization, such as the third and fifth class of the Fashion-MNIST dataset. This result implies that the normalization process that is preceded in the amplitude encoding can deteriorate the classification performance.
\begin{table}[ht]
\begin{tabular}{ccccc}
\hline
\multicolumn{2}{c}{}                                                                                                                   & \multicolumn{3}{c}{Average AUC (\%)}                                                                                     \\ \hline
\multicolumn{1}{l}{Dataset}                                                   & \begin{tabular}[c]{@{}c@{}}Normal\\ Class\end{tabular} & Amplitude     & \begin{tabular}[c]{@{}c@{}}Kernel\\ PCA\end{tabular} & \begin{tabular}[c]{@{}c@{}}OC-\\ SVM\end{tabular} \\ \hline
\multirow{10}{*}{\begin{tabular}[c]{@{}c@{}}Handwritten\\ Digit\end{tabular}} & \multicolumn{1}{c|}{0}                                 & $99.9\pm0.1$  & 99.9                                                 & 99.5                                              \\
                                                                              & \multicolumn{1}{c|}{1}                                 & $97.2\pm 0.8$ & 98.6                                                 & 80.0                                              \\
                                                                              & \multicolumn{1}{c|}{2}                                 & $97.6\pm 0.9$ & 99.3                                                 & 95.4                                              \\
                                                                              & \multicolumn{1}{c|}{3}                                 & $98.1\pm 0.3$ & 98.5                                                 & 97.7                                              \\
                                                                              & \multicolumn{1}{c|}{4}                                 & $98.0\pm 0.6$ & 97.0                                                 & 93.9                                              \\
                                                                              & \multicolumn{1}{c|}{5}                                 & $98.7\pm 0.5$ & 99.7                                                 & 95.3                                              \\
                                                                              & \multicolumn{1}{c|}{6}                                 & $99.8\pm 0.1$ & 99.1                                                 & 99.0                                              \\
                                                                              & \multicolumn{1}{c|}{7}                                 & $99.0\pm 0.5$ & 95.5                                                 & 98.7                                              \\
                                                                              & \multicolumn{1}{c|}{8}                                 & $92.6\pm 1.8$ & 95.2                                                 & 81.8                                              \\
                                                                              & \multicolumn{1}{c|}{9}                                 & $94.9\pm 1.3$ & 97.7                                                 & 87.9                                              \\ \hline
\multirow{10}{*}{\begin{tabular}[c]{@{}c@{}}Fashion-\\ MNIST\end{tabular}}    & \multicolumn{1}{c|}{0}                                 & $90.6\pm 0.6$ & 89.0                                                 & 87.2                                              \\
                                                                              & \multicolumn{1}{c|}{1}                                 & $98.2\pm 0.2$ & 98.5                                                 & 91.3                                              \\
                                                                              & \multicolumn{1}{c|}{2}                                 & $86.4\pm 0.6$ & 86.2                                                 & 87.8                                              \\
                                                                              & \multicolumn{1}{c|}{3}                                 & $91.9\pm 1.0$ & 91.0                                                 & 91.9                                              \\
                                                                              & \multicolumn{1}{c|}{4}                                 & $93.6\pm 0.2$ & 93.6                                                 & 93.6                                              \\
                                                                              & \multicolumn{1}{c|}{5}                                 & $64.0\pm 2.5$ & 75.3                                                 & 43.2                                              \\
                                                                              & \multicolumn{1}{c|}{6}                                 & $80.2\pm 0.4$ & 79.9                                                 & 80.4                                              \\
                                                                              & \multicolumn{1}{c|}{7}                                 & $97.8\pm 0.3$ & 98.1                                                 & 97.0                                              \\
                                                                              & \multicolumn{1}{c|}{8}                                 & $90.0\pm 1.0$ & 89.4                                                 & 80.8                                              \\
                                                                              & \multicolumn{1}{c|}{9}                                 & $99.0\pm 0.3$ & 98.9                                                 & 98.3                                             
\end{tabular}
\caption{Average AUCs in \% and one standard deviation (over 10 seeds) of the one-class classification on the handwritten digit and Fashion-MNIST dataset under normalization.}
\label{tab:norm_data}
\end{table}

\end{document}